\documentclass[pre,aps,twocolumn]{revtex4-1}
\usepackage{amsmath,natbib,amsfonts}
\usepackage{amssymb,color}
\usepackage{graphicx,lipsum}
\usepackage{bbold}
\newcommand{\bea}{\begin{eqnarray}}
\newcommand{\eea}{\end{eqnarray}}

\newcommand{\lan}{\langle}
\newcommand{\ran}{\rangle}

\begin{document}
\title{Two coupled, driven Ising spin systems working as an Engine}
\author{Debarshi Basu$^{1}$, Joydip Nandi$^{1}$, A. M. Jayannavar$^{2,3}$ and Rahul Marathe$^{1}$} 
\email{maratherahul@physics.iitd.ac.in, jayan@iopb.res.in}
\affiliation{$^1$Department of Physics, Indian Institute of Technology, Delhi, Hauz Khas 110016, New Delhi, India.\\
$^2$Institute of Physics, Sachivalaya Marg, Bhubaneshwar 751005, Odhisha, India.\\
$^3$Homi Bhabha National Institute, Training School Complex, Anushakti Nagar, Mumbai 400085, India. }

\date{\today}
\begin{abstract}
Miniaturized heat engines constitute a fascinating field of current research. Many  
theoretical as well as experimental studies are being conducted which involve colloidal particles 
in harmonic traps as well as bacterial baths acting like thermal baths. These systems are micron 
sized and are subjected to large thermal fluctuations. Hence for these systems
average thermodynamic quantities like work done, heat exchanged and efficiency loose meaning 
unless otherwise supported by their full probability distributions. Earlier studies on micro-engines
are concerned with applying Carnot or Stirling engine protocols to miniaturized systems, where
system undergoes typical two isothermal and two adiabatic changes. 
Unlike these models we study a prototype system of two classical 
Ising spins driven by time dependent, phase different, external magnetic fields. 
These spins are {\it simultaneously} in contact with two heat reservoirs at 
different temperatures for the full duration of the driving protocol. Performance of the model 
as an engine or a refrigerator depends only on a single parameter namely the phase between 
two external drivings. We study this system in terms of fluctuations in efficiency and coefficient 
of performance (COP). We find full distributions of these quantities numerically and  
study the tails of these distributions. We also study reliability of the engine. We find the
fluctuations dominate mean values of efficiency and COP, and their probability distributions are
broad with power law tails.      
\end{abstract}
\maketitle

\noindent{\it Introduction:} After Feynman's theoretical construction of his famous Ratchet and Pawl 
machine in \cite{Feynman66}, due to advancement in nano science, it is now possible to realize 
miniaturized engines experimentally \cite{exp1, exp2, Edgar16, Sood16}. 
Many of the experiments are based on theoretical predictions namely the fluctuation theorems which 
put bounds on thermodynamic quantities of interests like efficiency of the engines
\cite{Bustman05, AMJ12}. For thermodynamic engines like Carnot or Stirling 
the fluctuations are usually ignored and most of the physics is obtained from average values of 
work and heat \cite{Callen}. These notions however fail in case of microscopic engines. 
Micro-engines behave differently and the main reason behind this odd behavior are the {\it loud} 
thermal fluctuations. These thermal fluctuations cause energy exchanges of the order of $k_BT$, where 
$k_B$ is the Boltzmann constant and $T$ the ambient temperature. For small systems one can thus not 
just rely on mean values of work and heat or as a matter of fact any thermodynamic quantity, 
but one has to look at full probability distributions. To deal with such systems one needs 
to use the framework of stochastic thermodynamics \cite{Sekimoto97, Seki98, SekiBook}. 
Many studies on such small scale engines have shown that fluctuations in thermodynamic 
quantities dominate over mean values even in the quasistatic limit 
\cite{AMJ16_1, AMJ16, Arun14, Tu14, Holu14, Udo12, Broeck06, AMJ96}. 
Many studies have also looked at full distributions of efficiency \cite{AMJ16, Arun14}
and also the large deviation functions \cite{Broeck14}. Models with feedback control both 
instantaneous and delayed have also been investigated \cite{RSM16, Qian02, Qian04, Rosinberg15}. 
Most of the earlier studies both theoretical and experimental were based on applying the 
thermodynamic engine protocols like Carnot or Stirling to a colloidal particle placed in an 
harmonic trap. The trap strength is then modified time dependently to mimic isothermal expansion, 
compression and adiabatic expansion, compression steps \cite{exp1, exp2, AMJ16, Arun14}. We would 
like to point out that there are in fact no detailed studies which deal with fluctuations of 
thermodynamic quantities for externally driven systems which are 
simultaneously in contact with several heat baths. These systems show many novel features not seen 
in earlier studied models. 

\begin{figure}
\centering
\includegraphics[width=.8\columnwidth]{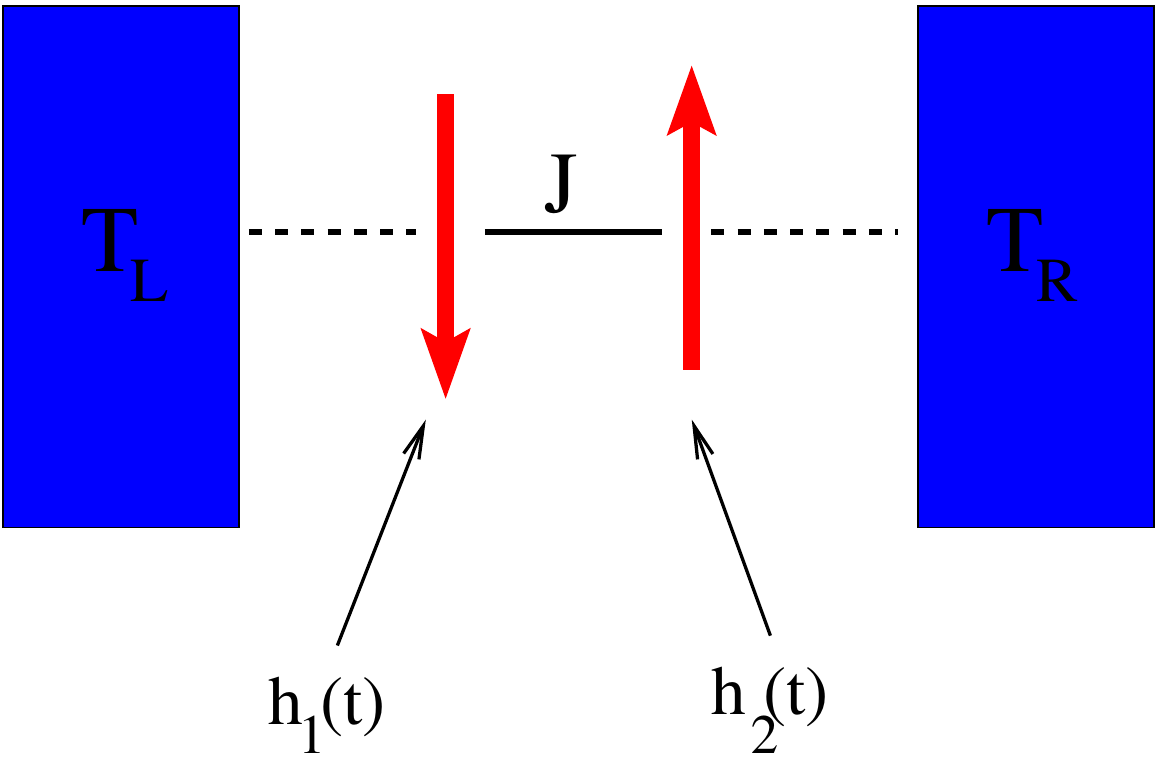}
\caption{(Color online) Cartoon of the model discussed in the text.}
\label{model}
\end{figure}

\begin{figure*}
\centering
\includegraphics[width=.8\columnwidth]{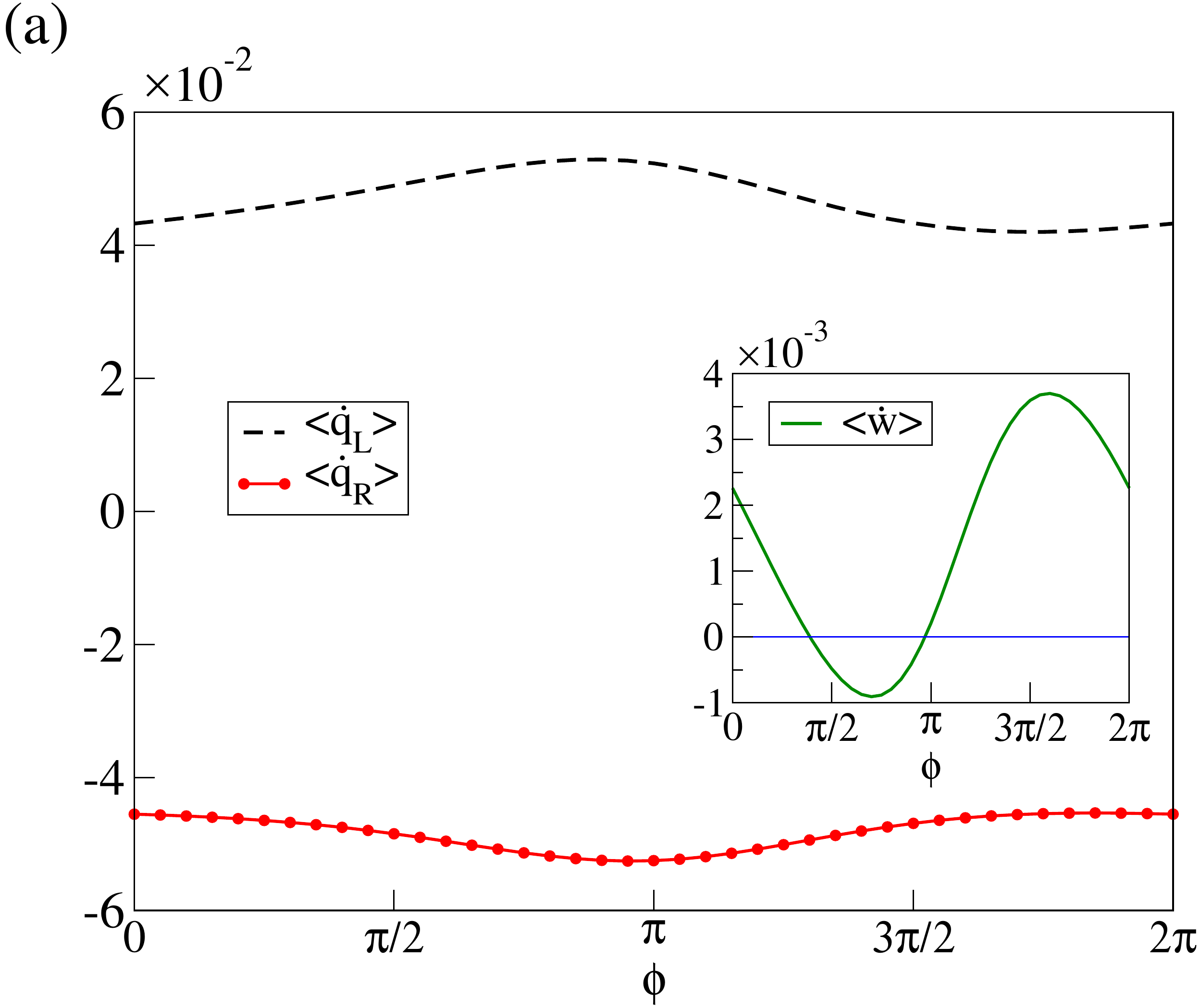}
\hspace{1.cm}
\includegraphics[width=.87\columnwidth]{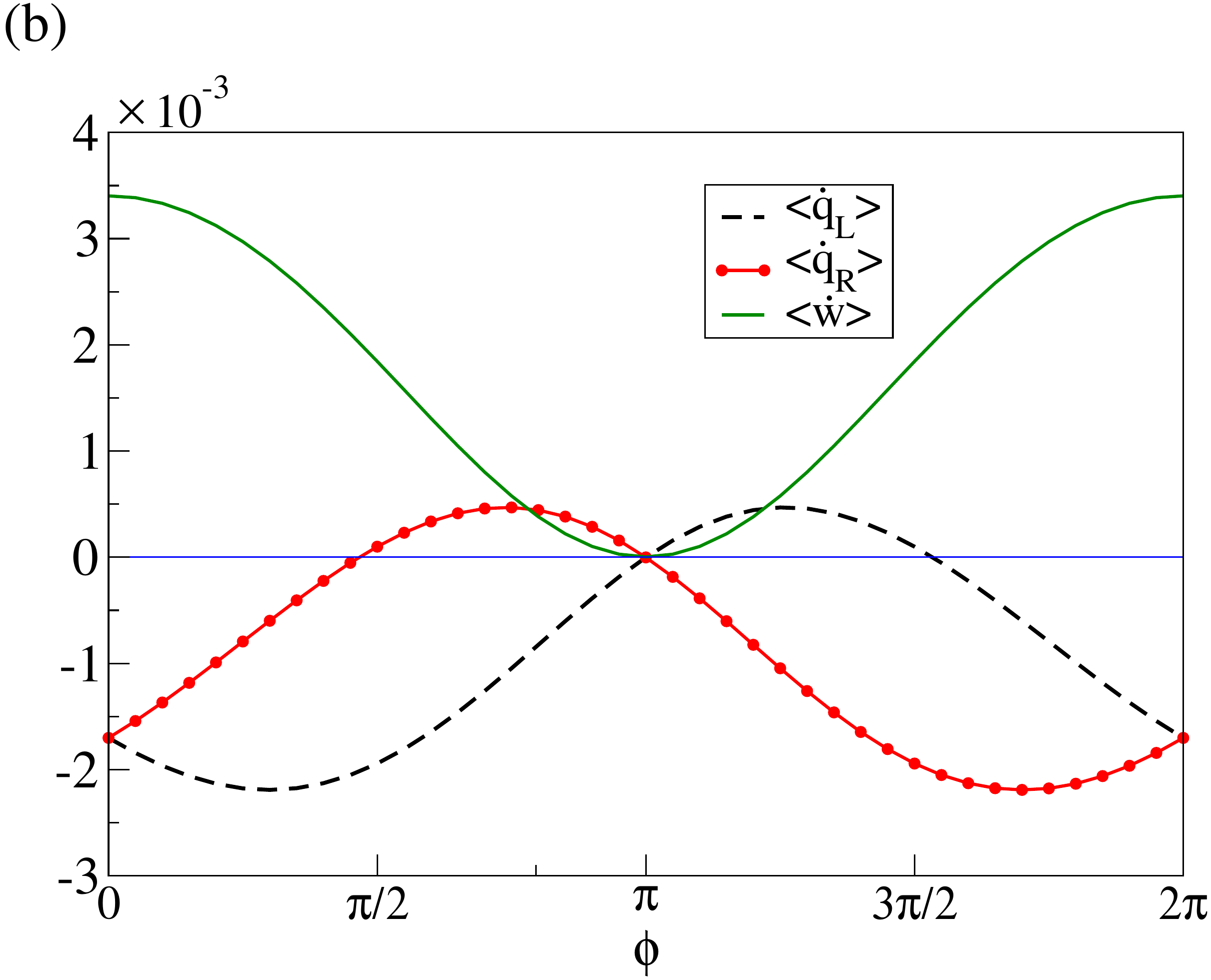}
\caption{(Color online) $(a)$ Engine mode of operation. $\lan\dot{q}_L\ran>0$, $\lan\dot{q}_R\ran <0$ and $\lan\dot{w}\ran<0$ for certain values of
the phase $\phi$. Parameter values are $h_0=0.25$, $\tau=190$, $T_L=1.0$, $T_R=0.1$. At $\phi=0.7\pi$
maximum work is extracted from the system (inset). In all the results discussed further these parameters
are considered to be optimal for engine mode of operation. $(b)$ Pump/refrigerator mode of operation. Parameter values are $h_0=0.25$, $\tau=225$, 
$T_L=0.5$, $T_R=0.5$. See for example at $\phi=0.7\pi$ we have $\lan\dot{q}_R\ran > 0$, $\lan\dot{w}\ran > 0$, and $\lan\dot{q}_L\ran < 0$ 
implying heat is taken from the right bath, work is done on the system and heat is dissipated 
into the left bath. In all the results discussed further these parameters are considered to be 
optimal for pump/refrigerator mode of operation. Zero line is just a guide for the eyes. 
See \cite{RSM07} for details.}
\label{fig1}
\vspace{-3mm}
\end{figure*}

In this work we have studied a model of classical heat engine and a pump where two Ising spins
are independently kept in contact with two heat baths at different temperatures. These spins
are externally driven by time dependent magnetic fields with a phase difference \cite{RSM07},
see Fig. \ref{model}. During full driving protocol system is never isolated from the heat baths.
Interestingly the phase difference is the only parameter which decides whether system works as a 
heat engine or a refrigerator. Performance of this model in terms of average heat currents has been
studied in \cite{RSM07}. In this paper we analyze this model in terms of following:
\begin{enumerate}
\item Rich features this model exhibits in phase diagrams of engine and pump performance. 
\item Fluctuations in efficiency, COP and their probability distributions, including power law tails. 
\item Behavior of work, heat, efficiency and power in quasistatic limit. 
\item Reliability of the model to work either as an engine or a refrigerator. 
\end{enumerate}

\noindent{\it Model:} We consider a model of two classical Ising spins with interaction energy 
$J$, driven by time dependent external magnetic fields  
$h_1(t)=h_0 \cos(\omega t)$ and $h_2(t)=h_0 \cos(\omega t+\phi)$, where $\phi$ is the phase 
difference and $\omega$ the driving frequency, as shown in Fig. \ref{model}. The Hamiltonian 
for this system is written as:
\bea
\mathcal{H}=-J \sigma_1\sigma_2 -h_1(t)\sigma_1-h_2(t)\sigma_2,~~~\sigma_{1,2} =\pm 1.
\eea
Left and right spins are in contact with heat baths at temperature $T_L$ and $T_R$ respectively.
Interaction of spins with the respective heat baths is modeled by Glauber dynamics \cite{Glauber63}.
We define heat currents coming from left(right) baths $\dot{Q}_L$($\dot{Q}_R$) and
work done on left(right) spin $\dot{W}_L$($\dot{W}_R$) to be positive. The total work done is
nothing but $\dot{W}=\dot{W}_L+\dot{W}_R$. If $P(\sigma_1,\sigma_2, t)$ represents the probability
to have spins in state $\{\sigma_1,\sigma_2\}$ at time $t$ then the heat exchange rates can be
written as:

\bea
\dot{Q}_L &=&\sum_{\sigma_1, \sigma_2}~ P(\sigma_1,\sigma_2, t)~ r_{\sigma_1, \sigma_2}^L 
~\Delta E_1(\sigma_1, \sigma_2) \nonumber \\
\dot{Q}_R &=&\sum_{\sigma_1, \sigma_2} P(\sigma_1,\sigma_2, t)~ r_{\sigma_1, \sigma_2}^R 
~\Delta E_2(\sigma_1, \sigma_2) \nonumber \\ 
\dot{W}_L &=& -\langle \sigma_1 \rangle~ \dot{h}_1(t) = -\dot{h}_1(t) \sum_{\sigma_1, \sigma_2} \sigma_1~ 
P(\sigma_1,\sigma_2, t) \nonumber \\ 
\dot{W}_R &=& -\langle \sigma_2 \rangle~ \dot{h}_2(t) = -\dot{h}_2(t) \sum_{\sigma_1, \sigma_2} \sigma_2~ 
P(\sigma_1,\sigma_2, t),
\label{QWrates}
\eea
where the modified Glauber spin flip rates to compensate for two heat reservoirs are given by:
\bea
r_{\sigma_1, \sigma_2}^{L,R} = r (1-\gamma_{L,R} \sigma_1\sigma_{2}) (1-\delta_{L,R} \sigma_{1,2}),
\eea
with:
\bea
\gamma_{L,R} = \tanh(J/k_BT_{L,R}), \nonumber \\
\delta_{L,R} = \tanh(h_{1,2}/k_BT_{L,R}),
\eea
where $r$ is a rate constant. The energy changes associated with left or right spin flips 
are given by: 
\bea
\Delta E_1 = 2(J\sigma_1\sigma_{2} + h_1(t) \sigma_1),\nonumber \\
\Delta E_2 = 2(J\sigma_1\sigma_{2} + h_2(t) \sigma_2),
\eea
Expressions in Eq.~(\ref{QWrates}) can be easily obtained 
from the master equation satisfied by $P(\sigma_1,\sigma_2, t)$, see \cite{RSM07} for details. 
It is easy to show that the average energy 
$U=\langle \mathcal{H}\rangle=\sum_{\sigma_1,\sigma_2}\mathcal{H}(\sigma_1,\sigma_2)~
P(\sigma_1,\sigma_2, t)$ and $\dot{U}=\dot{Q}_L+\dot{Q}_R+\dot{W}_L+\dot{W}_R$ from above expressions.
Since external driving is time dependent, after a transient period probability 
$P(\sigma_1,\sigma_2,t)$ attains a time periodic state which is independent of the initial 
state. We also define time averaged heat and work currents namely:
\bea
\lan\dot{q}_{L,R}\ran =1/\tau \int_0^\tau \dot{Q}_{L,R}~dt \nonumber \\
\lan \dot{w}\ran=1/\tau\int_0^{\tau} \dot{W}~dt, 
\eea
where $\tau=2\pi/\omega$ is the time period of the external driving. 

\begin{figure*}[tbhp]
\centering
\includegraphics[width=19cm, height=8cm, angle=0]{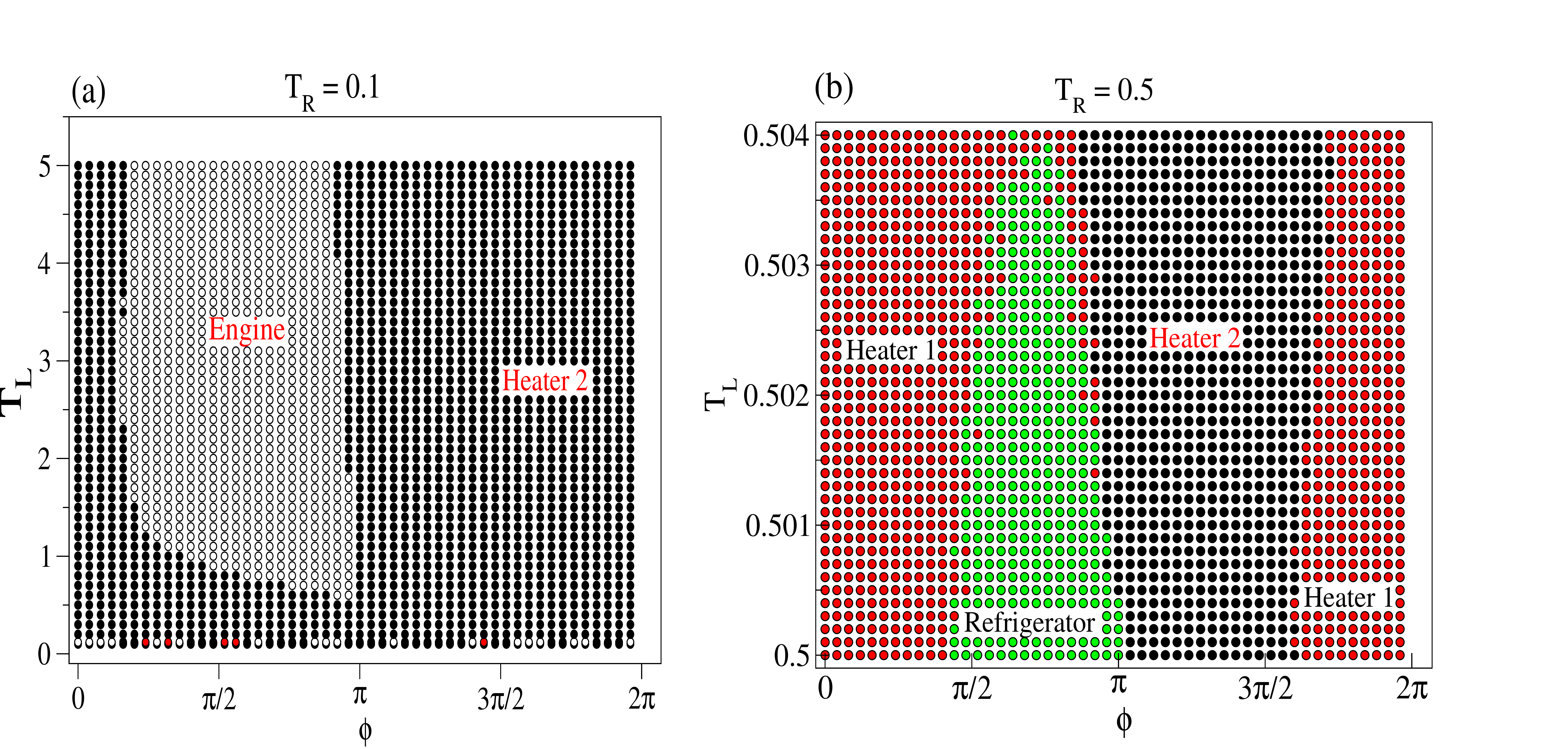}
\caption{(Color online) $(a)$ Shows the phase diagram as a function of the phase difference $\phi$ 
for engine mode of operation. Black solid circles indicate Heater $2$ mode and open black circles 
indicate Engine mode. $(b)$ Phase diagram for the refrigerator mode. Solid red circles Heater $1$ 
$($for $0\leq \phi < \pi/2$ and $3\pi/2 < \phi \leq 2\pi $ $)$ and 
solid green circles Refrigerator $($for $\pi/2 \leq \phi < \pi$ $)$, solid black circles 
Heater $2$ mode $($ for $\pi \leq \phi < 3\pi/2$ $)$. Different modes are also indicated 
in the figure.}
\label{phiphase}
\vspace{-3mm}
\end{figure*}

Once these definitions are set, for $T_L \geq T_R$, we define stochastic efficiency $\epsilon$ and 
stochastic coefficient of performance (COP) $\eta$ as:
\bea
\epsilon = \frac{\dot{w}}{-\dot{q}_L}\ ,~~ \eta =  \frac{\dot{q}_R}{\dot{w}}\ .
\eea
We note that due to large thermal fluctuations two efficiencies 
\[ \bar{\epsilon} = \frac{\langle\dot{w}\rangle}{\langle-\dot{q}_L\rangle}\
~\text{and}~ \langle \epsilon \rangle = \left \langle \frac{\dot{w}}{-\dot{q}_L}\ \right \rangle, \]
are in general not equal that is $\langle\epsilon \rangle\neq \bar{\epsilon}$ similarly 
$\langle\eta\rangle \neq \bar{\eta}$. For completeness we reproduce results from reference 
\cite{RSM07} to show how the phase $\phi$ and
time period $\tau$, determine the engine or pump behavior. In Fig. \ref{fig1} $(a)$, $(b)$, 
we plot $\lan \dot{w}\ran$, $\lan \dot{q}_L\ran$ and $\lan \dot{q}_R\ran$ as a function of the phase 
$\phi$ for engine and pump mode of operation respectively. In Fig. \ref{fig1} $(a)$, for all values 
of $\phi$ the heats $\lan \dot{q}_L \ran >0 $, $\lan \dot{q}_R\ran <0$ but for a narrow range 
$\pi/2 \leq \phi \leq \pi$ work done $\lan\dot{w}\ran < 0 $. 
In this narrow range work is extracted from the system hence the device works as an engine. 
It can be seen that for parameters $T_L=1.0$, $T_R=0.1$, $J=1.0$, $h_0=0.25$ 
and $\tau=190$, at $\phi=0.7\pi$ maximum work is extracted. We refer to 
this set of parameter values as optimal parameters for engine mode of operation throughout 
the manuscript. Similarly In Fig. \ref{fig1} $(b)$, for all values of $\phi$ work 
done $\lan\dot{w}\ran >0$ but the heats $\lan\dot{q}_L\ran$ and $\lan\dot{q}_R\ran$ take 
positive and negative values alternately. For a narrow strip $\pi/2 \leq \phi \leq \pi$,  
$\lan\dot{q}_L\ran < 0 $ and $\lan\dot{q}_R\ran > 0$ thus system works like a pump, transferring 
heat from right bath to the left. For parameters $T_L=0.5$, $T_R=0.5$, $J=1.0$, $h_0=0.25$ 
and $\tau=225$, at about $\phi=0.7\pi$ maximum pumping of heat happens. Thus we refer to these 
parameter values as optimal parameters for refrigerator/pump mode of operation throughout the 
manuscript. Similar results, as in Fig. \ref{fig1} $(b)$ are obtained if the 
right bath is slightly colder showing one can transfer heat from colder to hotter bath 
working as a refrigerator, see \cite{RSM07}. 

The average values of heats and work done can easily be obtained by solving the master equation 
numerically \cite{RSM07}. But to study fluctuations and distributions of these quantities we 
have to rely on Monte-Carlo simulations which we now describe.

\noindent{\it Simulations:} To study the dynamics of the system and for evaluating different 
heat currents, we perform Monte-Carlo simulations. 
We discretize the magnetic field sweep which consists of
$\sim 10^4$ time steps such that each time step $dt=\tau/10^4$ with $\tau=2\pi/\omega$ where $\tau$ 
is the time period of external driving. We also fix the Boltzmann constant $k_B=1$, the 
interaction energy $J=1.0$ and the rate constant $r=0.5$. 
Simulation follows usual Monte-Carlo steps in which first or second 
spin is chosen at random. At each discrete time step only one spin may flip. 
Since each spin is in contact with a separate heat bath, the spin flip rates themselves
can be used to evaluate flip probabilities by multiplying them with the time step $dt$.
In general flip rates need not be smaller than 1, thus we choose the rate constant $r$ such
that this problem does not arise \cite{Chvosta11}. 
At each step, if a spin flips, heat is exchanged between the left (right) spin and left 
(right) bath. We calculate these rates of heat exchange, the rate of work done on the first and 
second spin in the steady state, over one time period.

\begin{figure*}[tbhp]
\centering
\includegraphics[width=7.5cm, height=6.5cm, angle=0]{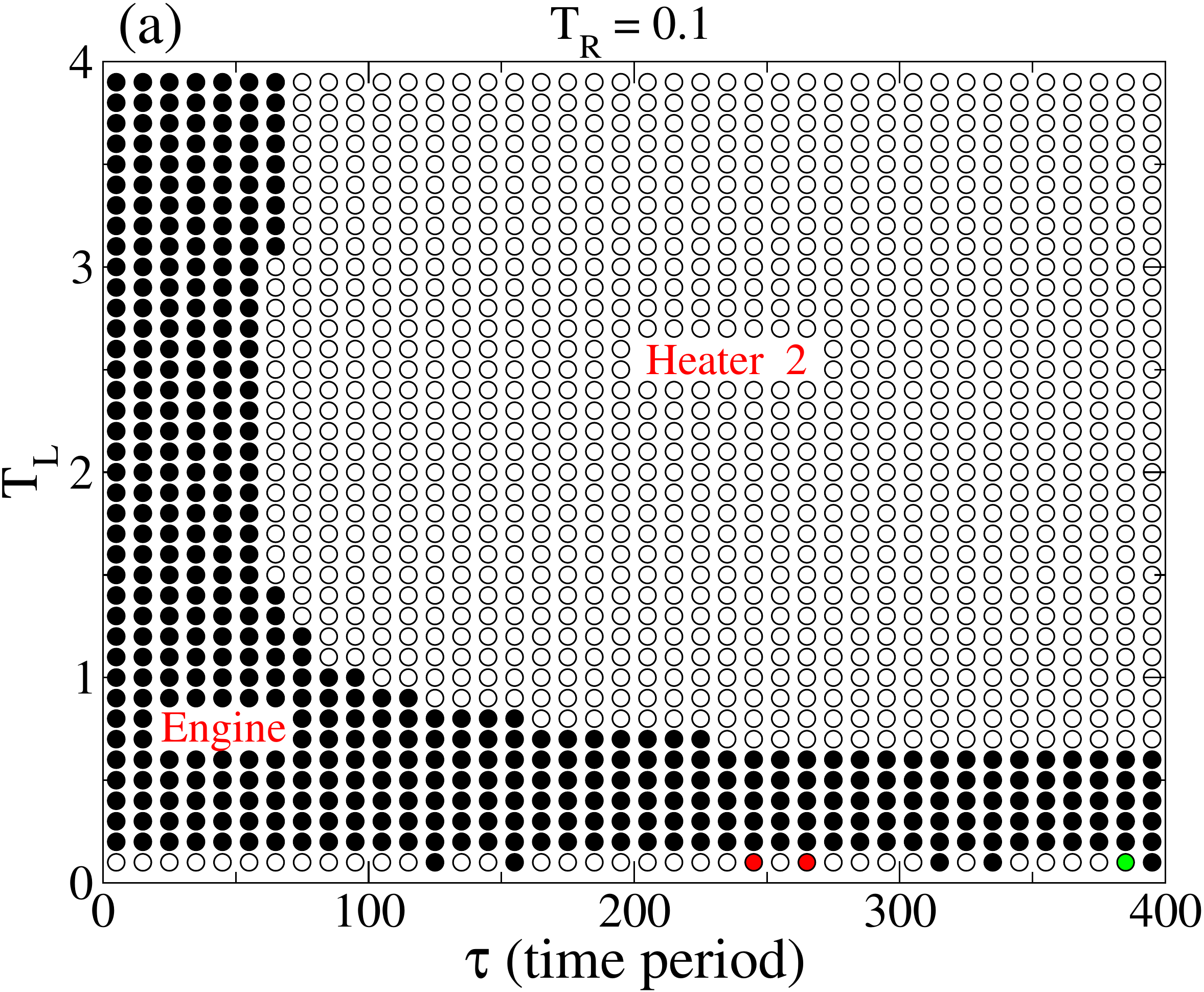}
\hspace{1cm}
 \includegraphics[width=7.5cm, height=6.5cm, angle=0]{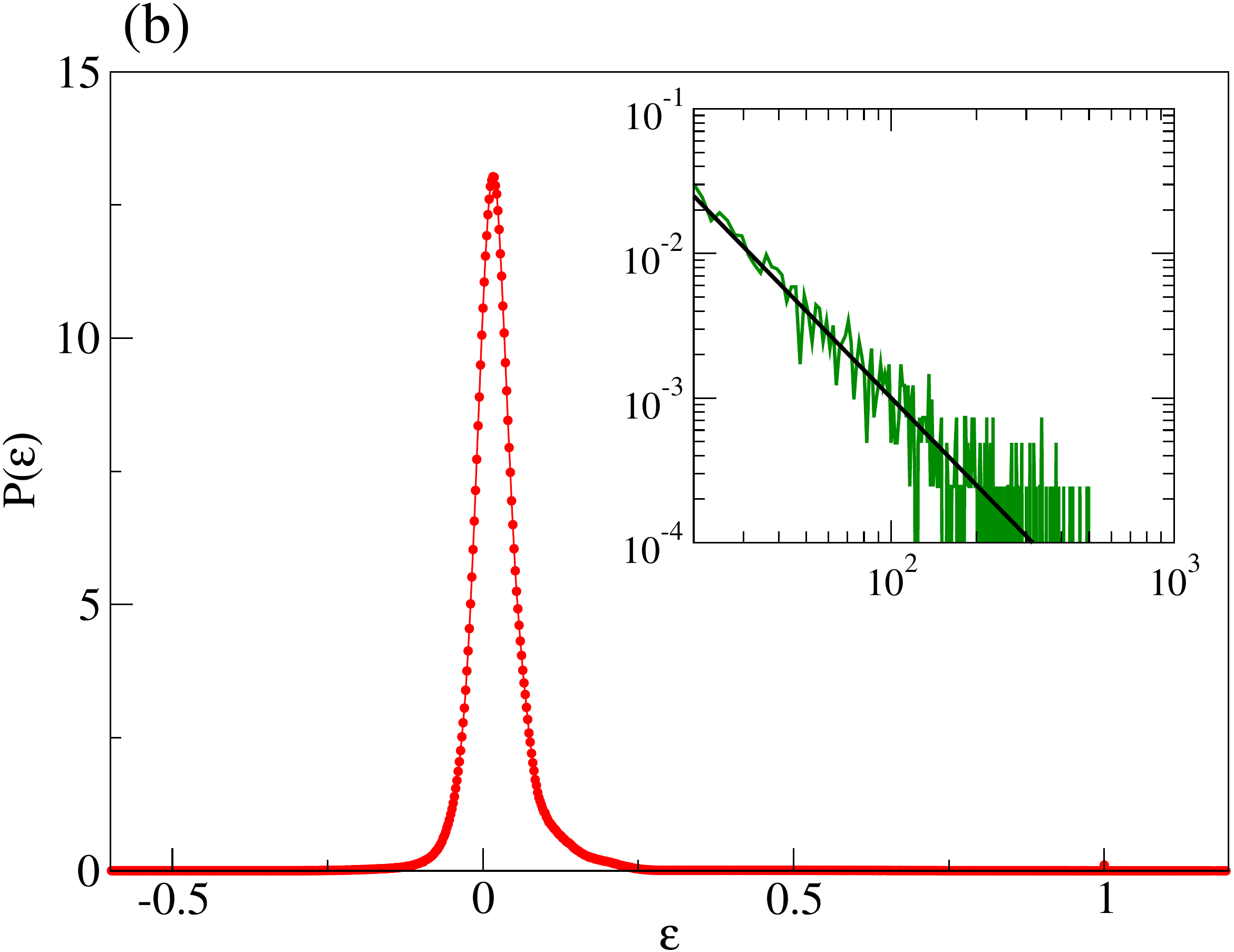}
\vspace{1cm}
\hspace{-2cm}
 \includegraphics[width=8cm, height=6.5cm, angle=0]{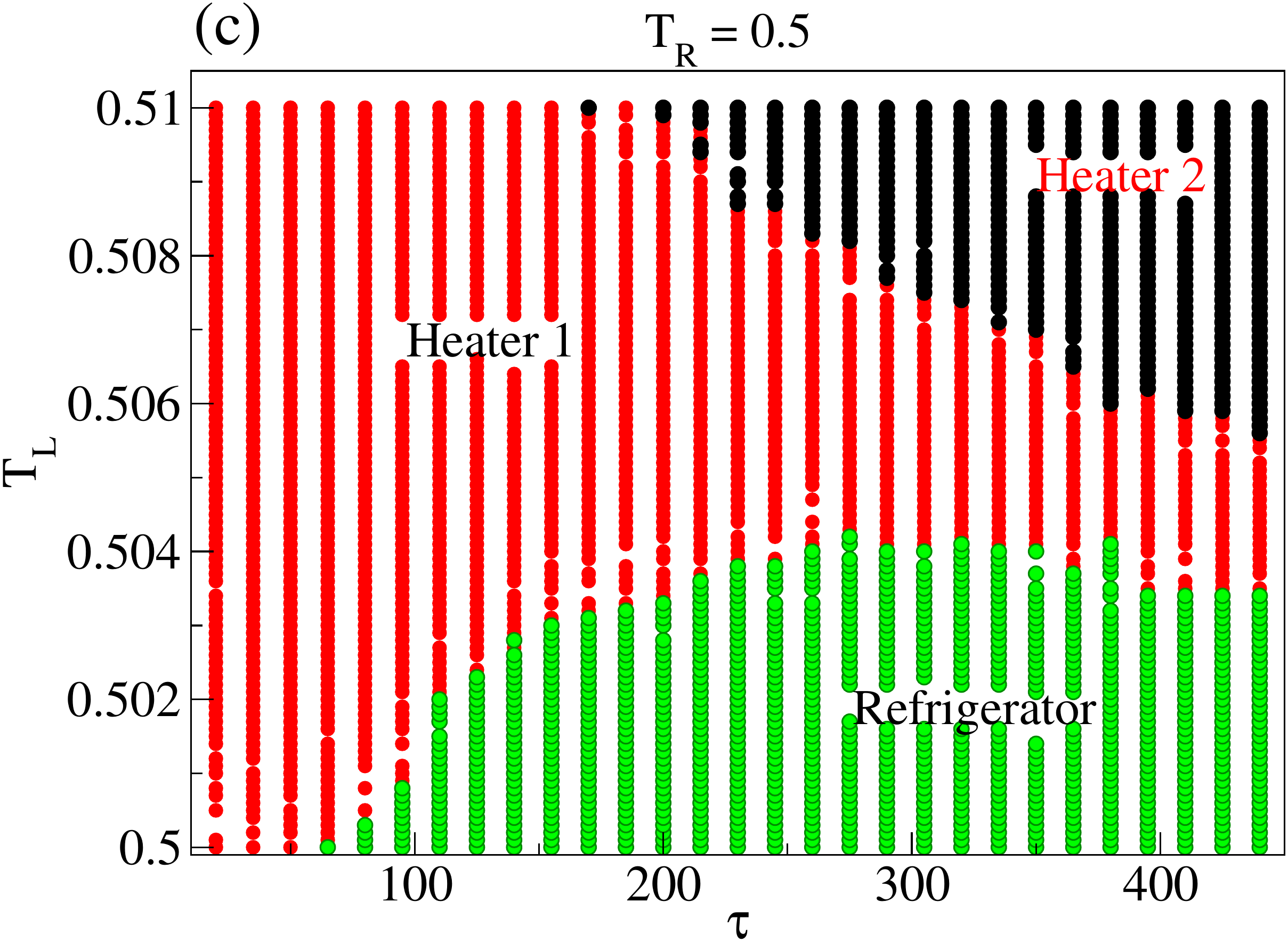}
\hspace{1cm}
 \includegraphics[width=8cm, height=6.5cm, angle=0]{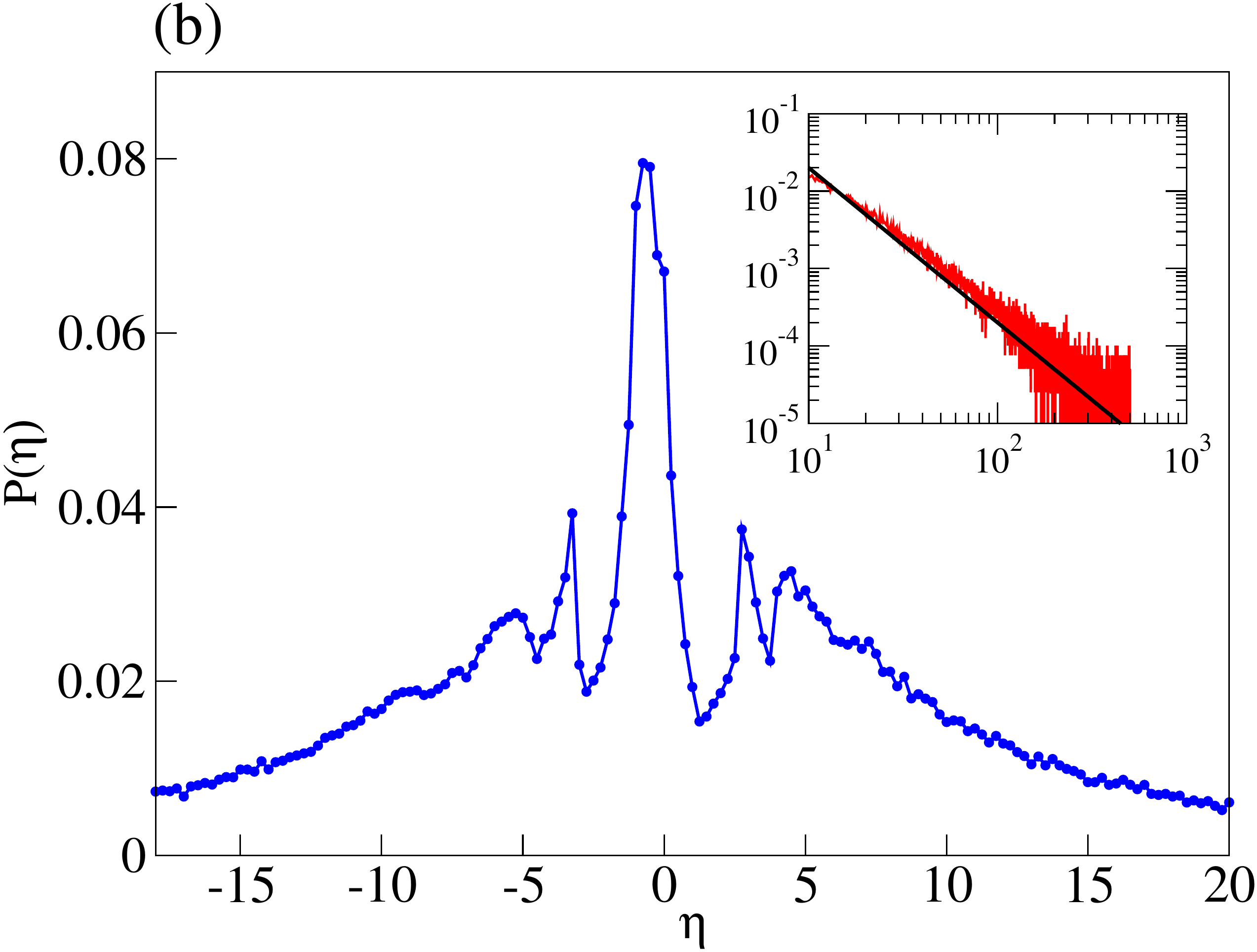}
   \caption{(Color online) $(a)$ Phase diagram for Engine mode of operation. 
Here $T_R=0.1$, $\phi=0.7\pi$, $h_0=0.25$
and $J=1.0$. Black solid circles represent engine operation while black open circles heater $2$ 
operation. These modes are also indicated in the figure. $(b)$ Distribution $P(\epsilon)$ 
of efficiency $\epsilon$ in the engine mode of 
operation, parameters used are $T_L=1.0$, $T_R=0.1$, $\phi=0.7\pi$, $h_0=0.25$ and $\tau=190$. Inset 
shows the tail part of the distribution for $\tau=10$. Tail of the distribution can be fitted to a power law $a\epsilon^{-\alpha}$ with exponent close to $2$ (solid black line).
$(c)$ Phase diagram for refrigerator mode of operation. Parameters are $T_R=0.5$, $\phi=0.7\pi$, $h_0=0.25$. Red portion $($ middle portion $)$ shows Heater $1$ operation, while Black portion represents Heater $2$ operation $($ upper right part$)$, and green Refrigerator operation $($ bottom right part $)$ is indicated in the figure. For Refrigerator mode of operation one requires the temperature difference between $T_L$ and $T_R$ to be small. $(d)$ Distribution $P(\eta)$ of coefficient of performance $\eta$ in the Refrigerator mode of operation. Parameters used are $T_L=0.5$, $T_R=0.5$, 
$\phi=0.7\pi$, $h_0=0.25$ and $\tau=225$. Inset shows the tail part of the distribution. Tail of the distribution can again be fitted to a power law $b\eta^{-\alpha}$ with exponent close to $2$ (solid black line).}
\label{phaseengref}
\end{figure*}
\begin{figure*}[tbhp]
\centering
\includegraphics[width=.65\columnwidth]{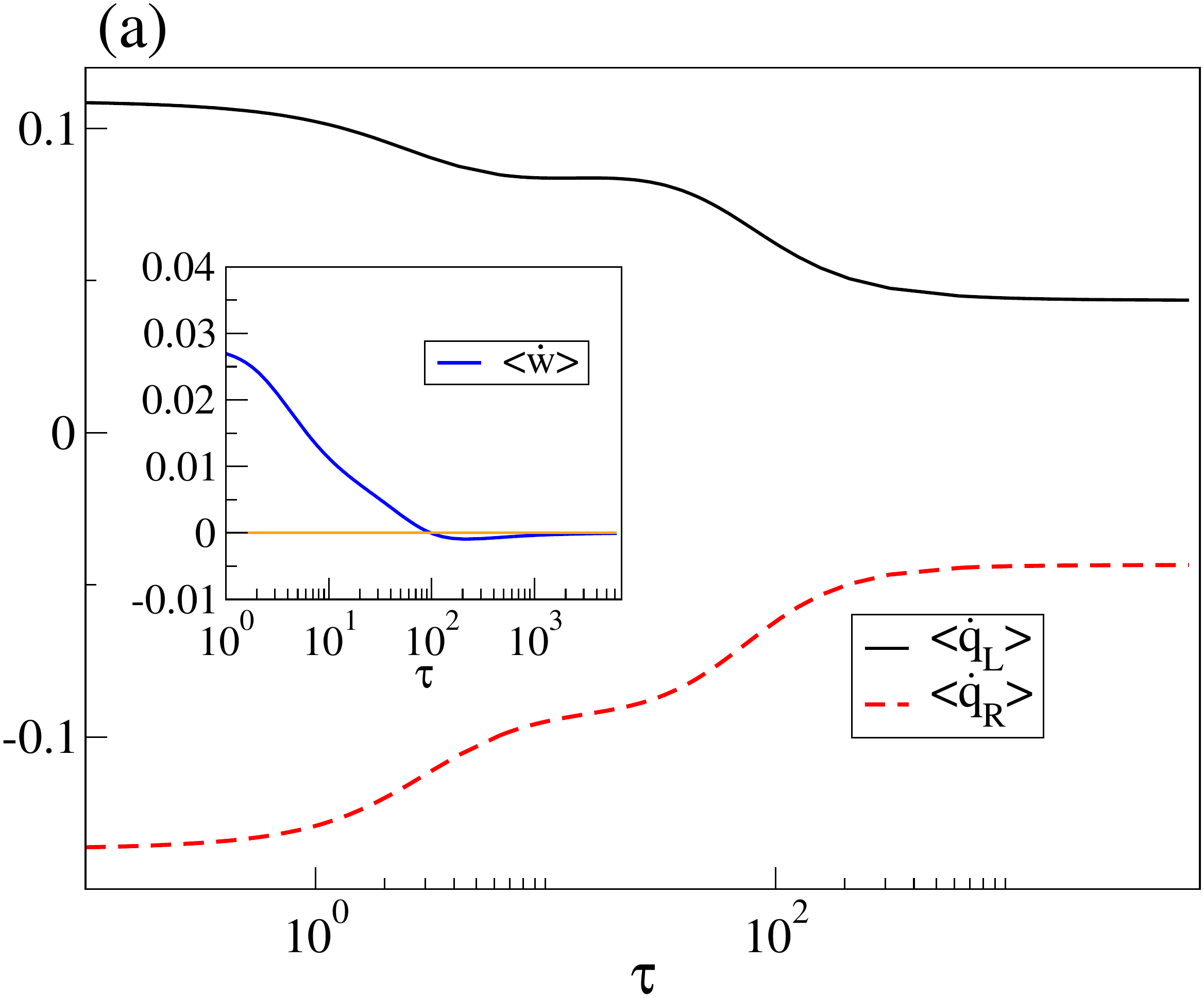}
\hspace{0.15cm}
\includegraphics[width=.68\columnwidth]{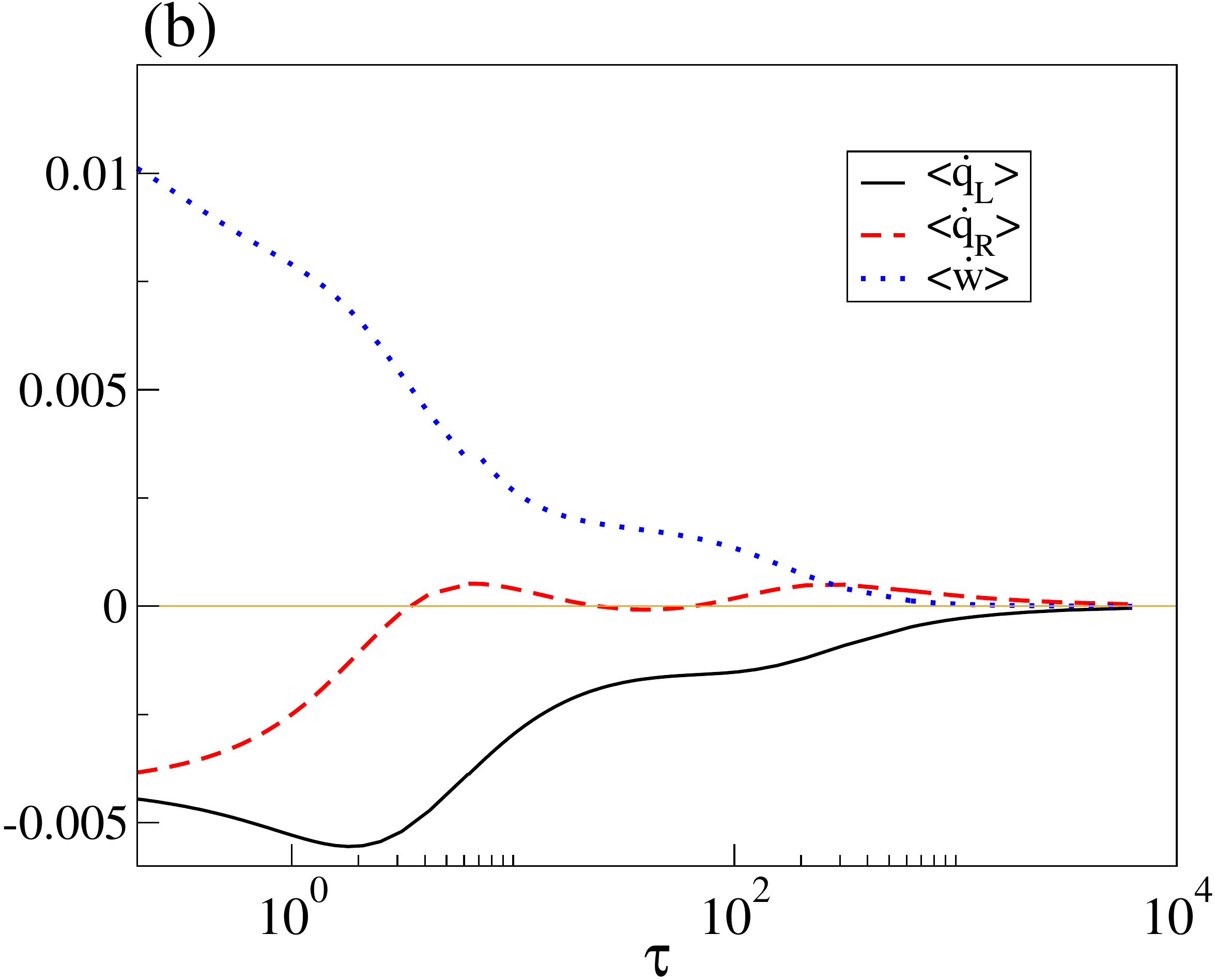}
\hspace{0.15cm}
\includegraphics[width=.66\columnwidth]{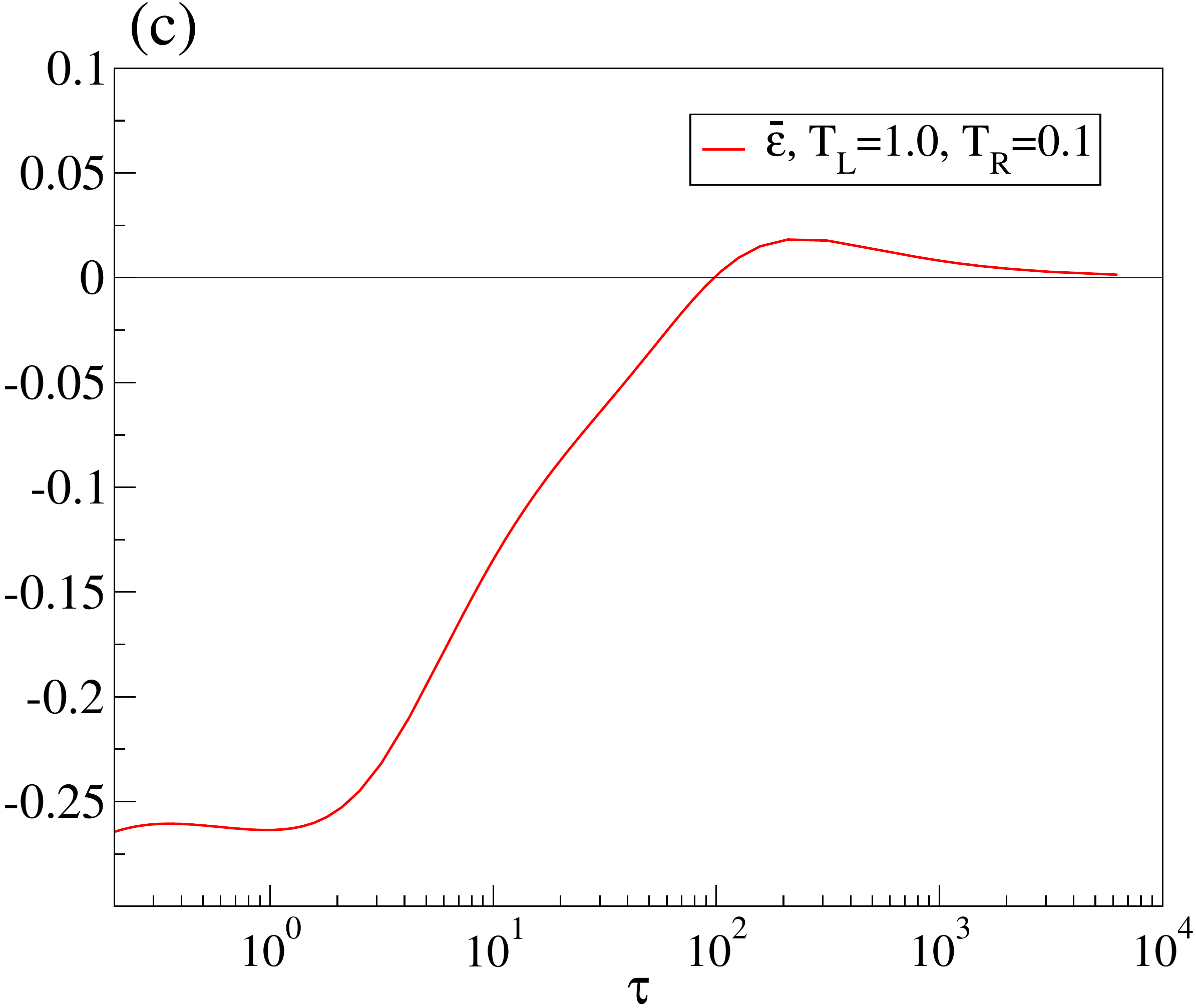}
\caption{(Color online) Plot of $\lan\dot{q}_L\ran$, $\lan\dot{q}_R\ran$, $\lan\dot{w}\ran$ as a function of time period of external driving $\tau$ in different modes of operation. 
Here $h_0=0.25$, $\phi=0.7\pi$. $(a)$ Engine mode, $T_L=1.0$, $T_R=0.1$. $(b)$ Refrigerator Mode, $T_L=0.5$, $T_R=0.5$. $(c)$ Efficiency $\bar{\epsilon}$ as a function of 
time period of external driving $\tau$ in engine mode, $T_L=1.0$, $T_R=0.1$. 
Zero line is just a guide for the eyes.}
\label{qwepvstau}
\end{figure*}

\begin{figure}[t]
\centering
\includegraphics[width=.9\columnwidth]{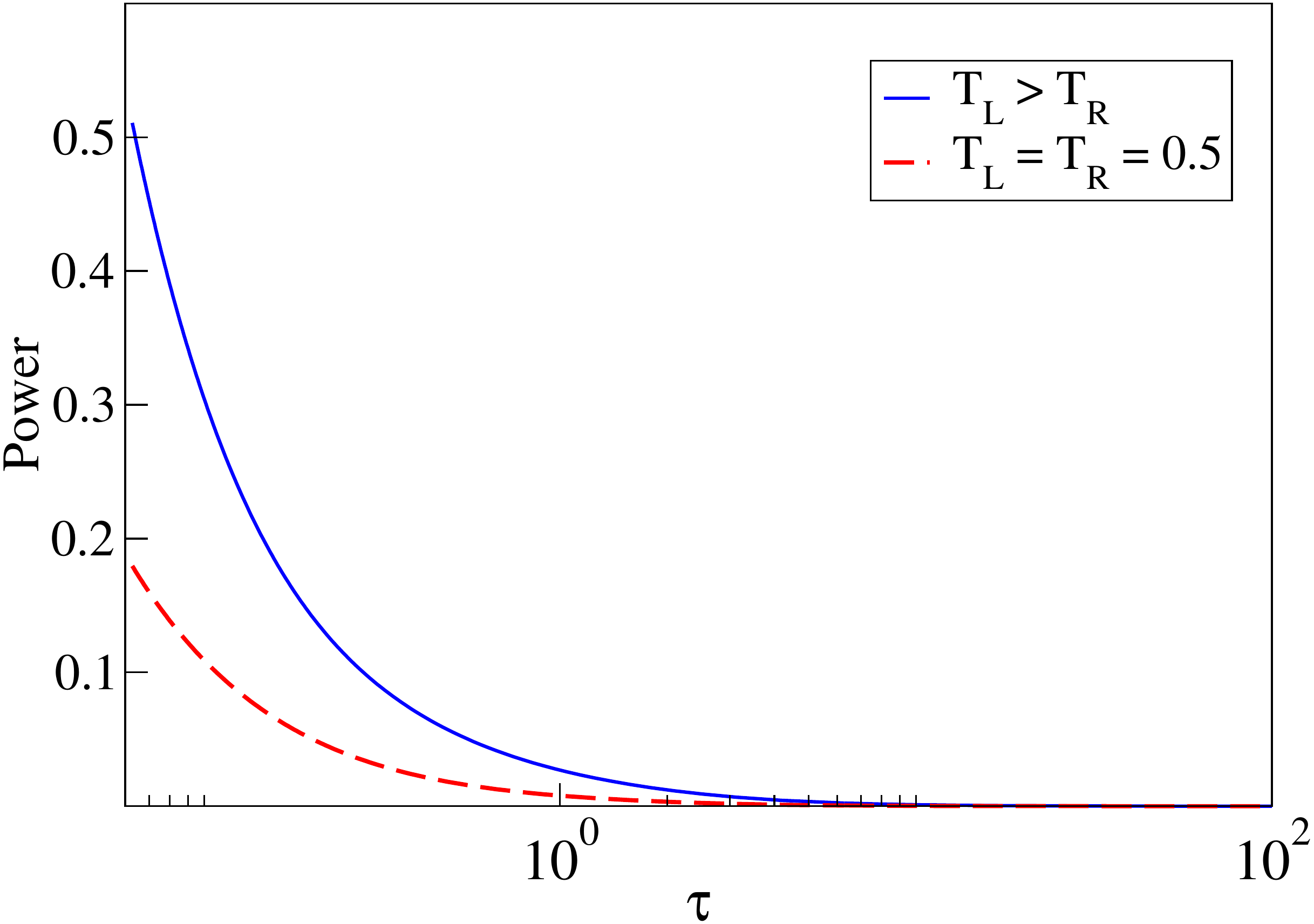}
\caption{(Color online) Plot of Power generated namely $\lan\dot{w}\ran/\tau$ as a function 
of time period 
of external driving $\tau$. For two sets of parameters $h_0=0.25$, $\phi=0.7\pi$, $T_L=1.0$, 
$T_R=0.1$ and $T_L=0.5$, $T_R=0.5$.}
\label{powvst}
\end{figure}

For our systems, there are four thermodynamically possible machines which are Engine, Heaters 1 
and 2, and Refrigerator \cite{AMJ16}. The actual mode of operation is determined by signs 
of heat exchanges $\lan\dot{q}_L\ran$, $\lan\dot{q}_R\ran$ and the total work done $\lan\dot{w}\ran$.
For $T_L\geq T_R$ these modes of operation are described as:
\begin{enumerate}
\item Engine mode: $\lan\dot{q}_L\ran > 0$, $\lan\dot{q}_R\ran < 0$, $\lan\dot{w}\ran < 0$, implying 
heat flows from left bath into the system, which is used by the working substance to do work on 
the external agent and remaining heat is dissipated into the right bath. 
\item Heater $1$ Mode: $\lan\dot{q}_L\ran < 0$, $\lan\dot{q}_R\ran < 0$, $\lan\dot{w}\ran > 0$. 
In this case external agent delivers large amount of heat in form of work into system and 
this heat is then dissipated in both left and right reservoirs. 
\item Heater $2$ Mode: $\lan\dot{q}_L\ran > 0$, $\lan\dot{q}_R\ran < 0$, $\lan\dot{w}\ran > 0$ 
heat flows from the left bath, as well as work is done on the system, hence a large amount 
of heat is dissipated in the right bath. 
\item Refrigerator Mode: $\lan\dot{q}_L\ran < 0$, $\lan\dot{q}_R\ran > 0$, $\lan\dot{w}\ran > 0$, 
heat is taken from the right bath which is at a slightly lower temperature than the left bath, 
work is done on the system and this results in transfer of heat to the left bath.
\end{enumerate}
In our model the phase difference $\phi$ and the time period $\tau$ alone can determine 
different modes of operations as can be seen from Fig. \ref{fig1}$(a)$, $(b)$ and 
Fig. \ref{phiphase}$(a)$, $(b)$.

We are also interested in studying fluctuations in heat exchanged and work done as well as to study 
how sensitive is the performance of the model in engine and pump/refrigerator mode, on the 
optimal parameter values described above. Hence we construct phase diagram for both modes of 
operations as a function of the phase $\phi$ and the temperature $T_L$ keeping $T_R=0.1$ for 
Engine mode and $T_R=0.5$ for pump mode of operation. These phase diagrams are shown in 
Fig. \ref{phiphase} $(a)$ and $(b)$ respectively. Fig. \ref{phiphase} $(a)$ shows how Engine mode
of operation depends on the phase $\phi$ and the temperature $T_L$ for fixed $T_R=0.1$ and 
$\tau=190$. It has two distinct domains namely Engine and Heater $2$. For $\pi/4 < \phi < \pi$ 
and $T_L-T_R > 1$, Engine behavior is observed $(\lan\dot{q}_L\ran > 0$, $\lan\dot{q}_R\ran < 0$, 
$\lan\dot{w}\ran < 0)$. Other part of the diagram is dominated by Heater $2$ operation. 
In Fig. \ref{phiphase} $(b)$ we plot phase diagram for the Refrigerator mode of operation where 
$T_R=0.5$, $\tau=225$. It is equally dominated by Heater $1$, Refrigerator and Heater $2$ modes
with refrigerator mode occurring in a narrow strip between $\pi/2 < \phi < \pi$, and for very small 
temperature differences $T_L-T_R \leq 0.005$. 

We now examine how different modes of operations depend on different parameters in the model other
than the phase $\phi$. To this end we construct the phase diagram where we keep the phase 
$\phi=0.7\pi$, temperature $T_R=0.1$ fixed and vary $T_L$ for different time periods of driving 
$\tau$. This phase diagram is shown in Fig. \ref{phaseengref}$(a)$. We see that for small 
$\tau\sim50$, work done $\lan\dot{w}\ran < 0$ with $\lan\dot{q}_L\ran > 0$, $\lan\dot{q}_R\ran < 0$, 
system works as an Engine independent of the temperature difference $T_L-T_R$. For large $\tau>50$ 
engine behavior persists but only for the moderate temperature differences $T_L-T_R\sim 1$. Other 
part of the phase diagram is mainly dominated by the Heater $2$ mode of operation where 
$\lan\dot{q}_L\ran > 0$, $\lan\dot{q}_R\ran < 0$, $\lan\dot{w}\ran > 0$. 
After determining the phase diagram we choose optimal parameters and find the probability 
distribution of efficiency $P(\epsilon)$. This distribution is shown in Fig. \ref{phaseengref}$(b)$. 
We see that the distribution is quite broad and has long power law tails $($inset of 
Fig. \ref{phaseengref} $(b)$. We would like to point out that in the quasistatic limit 
$\tau >100$ distribution becomes more and more peaked and tails become shorter. But for 
small $\tau\sim10$ tails of the distribution are long with power law decay.

Similar to Engine mode of operation discussed above we also look at the pump/refrigerator mode. In this
case we keep phase $\phi=0.7\pi$, $T_R= 0.5$ fixed, and change $T_L$ for different values of the time 
period $\tau$. This phase diagram is presented in Fig. \ref{phaseengref}$(c)$. Refrigerator 
$(\lan\dot{q}_L\ran < 0$, $\lan\dot{q}_R\ran > 0$, $\lan\dot{w}\ran > 0)$, mode occurs in a thin band 
for $\tau \geq 100$ for temperature differences $T_L-T_R\sim 0.005$. Other regions of the phase 
diagram are namely dominated by Heater $1$ $(\lan\dot{q}_L\ran < 0$, $\lan\dot{q}_R\ran < 0$, 
$\lan\dot{w}\ran > 0)$, for $\tau <100$ and $T_L-T_R >0.005$. Heater $2$ mode 
$(\lan\dot{q}_L\ran > 0$, $\lan\dot{q}_R\ran < 0$, $\lan\dot{w}\ran > 0)$ appears for larger values 
of $\tau > 200$ and larger temperature differences. We also plot the distribution of COP $P(\eta)$ 
in Fig. \ref{phaseengref} $(d)$. We see that distribution is broad with many distinct minima and 
long power law tails $($ inset Fig. \ref{phaseengref}$(b))$ with exponent $\sim -2$.

We also look at the behavior of different average heat currents namely $\lan \dot{q}_L\ran$, 
$\lan\dot{q}_R\ran$, $\lan\dot{w}\ran$ as a function of the driving period
$\tau$. This is crucial in order to understand how this engine performs when compared to the 
Carnot engine. In Fig. \ref{qwepvstau}$(a)$ we plot these currents for the
engine mode, where as expected $\lan \dot{q}_L\ran > 0$, $\lan\dot{q}_R\ran < 0$ for all $\tau$ 
values and they saturate to some finite value in the quasistatic limit
$\tau \rightarrow \infty$. However work done is negative only for a short interval when 
$\tau\sim 100$ (inset  of Fig. \ref{qwepvstau}$(a)$). Fig. \ref{qwepvstau}$(b)$
shows the Refrigerator mode where behavior changes from Heater $1$ for $\tau\sim 10$ to Refrigerator 
$(\tau\sim 50)$ and then to Heater $1$ for $\tau\sim 100$. 
Refrigerator mode recurs for $\tau\sim 500$ before all heat currents vanish in the quasistatic limit. 
Lastly in Fig. \ref{qwepvstau}$(c)$ we plot average efficiency 
$\bar{\epsilon}$ as a function of $\tau$ where for $\tau<100$ system is in the Heater $2$ mode 
$(\lan \dot{q}_L\ran > 0$, $\lan\dot{q}_R\ran < 0 $, $\lan\dot{w}\ran > 0$ 
(see Fig. \ref{qwepvstau}$(a))$, reaches a maximum value $\bar{\epsilon}\sim 0.025$ at $\tau\sim 190$ 
and then vanishes as $\tau\rightarrow\infty$, in quasistatic 
limit. This is consistent with the fact that though $\lan\dot{q}_L\ran$ is finite at large $\tau$ 
$($see Fig. \ref{qwepvstau}$(a))$, work done actually approaches zero in the quasistatic limit
$($inset of \ref{qwepvstau}$(a))$. This behavior is absent in usual colloidal engines where 
efficiency actually approaches Carnot efficiency in the quasistatic limit, distinguishing our model 
from earlier models \cite{AMJ16, Arun14}. Finally in Fig. \ref{powvst} we plot Power 
$\lan\dot{w}\ran/\tau$, as a function of the time period $\tau$ for fixed $T_L$, $T_R$ and $\phi$. 
As expected, for $\tau\sim 1$ finite amount of power is generated but it approaches zero as 
$\tau$ is increased. 

\noindent{\it Conclusion:} 
To conclude we have studied a novel model of two classical Ising spin interacting simultaneously 
with two heat baths and driven by time dependent, phase different magnetic
fields. Unlike earlier models the working substance is in contact with heat baths for the
full duration of the driving protocol. 
We also found that the performance of the system as an 
engine or a pump is highly affected by thermal fluctuations. For usual heat engines 
e.g. colloidal particles in contact with multiple baths, one expects that the efficiency 
should approach Carnot limit $1-T_c/T_h$ in the quasistatic or under zero power generation limit
\cite{Edgar16}. Since our model is in contact with both heat baths simultaneously, 
efficiency never reaches the Carnot limit due to non-zero entropy production even 
in the quasistatic limit. This is consistent with the B\"uttiker-Landauer model 
\cite{Buttiker87, Landauer88, Kawai08, Benjamin14}. This non zero entropy production rate defined as 
$\lan \dot{S}\ran= \lan -(\dot{q}_L/T_L)-(\dot{q}_R/T_R) \ran $, can be seen
from Fig. \ref{qwepvstau} $(a)$, where $\lan \dot{q}_L\ran $, $\lan \dot{q}_R\ran$ are non zero
in the quasistatic case. In fact in our model the efficiency 
goes to zero as the time period $\tau\rightarrow\infty$ as seen Fig. \ref{qwepvstau}$(c)$. 
We also point out that when the limit of small temperature difference and small driving frequency 
is taken simultaneously, the efficiency still remains much smaller than the 
Carnot efficiency. Similarly the COP is much smaller than the Carnot bound for the same reason.
Reliability of the engine is an important technological issue. 
Here reliability implies for how many cycles out of the total cycles, 
over which the averages are calculated, the device actually performed as an engine. 
We found that for optimal parameters in engine mode of operation with $\tau=190$ the 
reliability was about $75\%$. It was also seen that for $\tau\sim 10$ 
reliability was about $35\%$. As the time period increased $(\tau\sim 2000)$ reliability almost 
reached $100\%$ showing similar behavior as that of macroscopic engines. COP also shows similar 
behavior. This again points to the fact that fluctuations largely affect the performance. 
One interesting issue would be to look for possible ways to optimize the power and efficiency, 
on which we are currently working. To quantify fluctuations more concretely, we also numerically 
obtained probability distribution functions for efficiency $P(\epsilon)$ and COP $P(\eta)$. 
We found distributions to be very broad with power law tails, with exponent $\sim -2$. 
This points to the fact that fluctuations about the mean are much larger than unity. Currently 
we are studying the possibilities of an optimal protocol to increase the reliability of
the engine which may be independent of the time period.

\noindent{\it Acknowledgements:} 
RM thanks DST, India for financial support. DB, JN, RM thank the IIT Delhi HPC facility 
for computational resources. AMJ also thanks DST, India for J. C. Bose National Fellowship. 
Authors thank Varsha Banerjee for careful reading of the manuscript. We also thank one of
the referees for making many valuable suggestions towards the improvement of this manuscript.
 

\end{document}